# Basic physics of solid materials

Cao, T. D.
*Department of physics, Nanjing University of Information Science & Technology, Nanjing 210044, China*

Since the cuprate materials behave many curious properties, the mechanism that causes high-temperature superconductivity in copper oxide materials (cuprates) has been a controversial topic. To explain these curious properties, we must look out the physics what these materials include. In this letter, it is presented that there are the basic physics (1)-(9) in solid materials. I divide all solid materials into four categories, explain some properties of them qualitatively, and suggest some ways to turn a solid material into a superconductor. Particularly, this letter discusses the differences between the p-type high-$T_c$ cuprate superconductors, the n-type superconductor $Nd_{2-x}Ce_xCuO_4$, the BCS-superconductor similar $MgB_2$, and heavy Fermions superconductors.

After having read many articles about experiments and theory in literatures, I present these possible physics: (1) competition between correlations–strong spin correlation elevates the pairing temperature and reduces the pairing correlation length while strong charge correlation reduces the pairing temperature and extends the superconducting correlation length; (2) two kinds of electron pairs–one is the strong correlations induced pairs and the other is the phonons induced pairs; (3) conditions appearing superconductivity–the superconductivity appears when $\xi_{pair} > \bar{r}$ and $\bar{\tau} > \bar{r}/c$ ($\xi_{pair}$ is the correlation length between pairs, $\bar{r}$ the mean distance between pairs, $\bar{\tau}$ the average life-span of pairs, and $c$ the light speed); (4) strong correlations induced pairs do not contribute to superconductivity until phonon induced pairs have contributed to superconductivity; (5) the relation between superconductivity and pseudogap[1]– superconducting gap is just the pseudogap at T=$T_c$ and the electron pairing is the precursor of superconductivity; (6) the relations between strong correlations and excitations–the strong spin correlation and the strong charge correlation weaken with the enhancements of their excitations; (7) if the pairing temperature from phonons $T^* \gg T_c$, the isotope effect index in superconducting gap is small[2]; (8) possible roles of phonons–they result in s-wave symmetry dominated pseudogaps, evident isotope effects, $T$-linear resistivity at $T > \Theta_D/5$, and so on; and (9) possible roles of strong correlations–they result in non-s wave symmetry dominated pseudogaps, weak isotope effects, and spin correlation lead to $d^2R/dT^2 > 0$ while charge correlation lead to $d^2R/dT^2 < 0$, and so on. These physics are not isolated one another, for example, the physics (4) can be derived by (1) and (3). Although no authors have presented one of these physics in a complete way in the past, some similar ideas appear at some articles.

Since the spin correlation decreases with increased doping while the charge correlation increases with increased doping for cuprates, we introduce the moderate length $L_0$ by defining $L_0 = L_{spin} = aL_{charge}$ at $x_{opt}$ and $T_c^{max}$. $a$ is a proportional coefficient, $x_{opt}$ is the optimal doping, and $T_c^{max}$ is the highest critical temperature. This is shown in Fig.1.

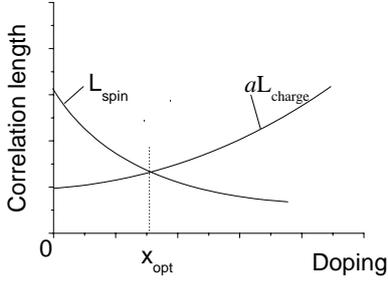

**Figure 1** The changes of correlation lengths with doping at the highest critical temperature of a cuprate superconductor.

Using $L_0$ is convenient for us to explain why the strong correlations dominated superconductors have various superconducting transition temperatures. On the basis of Fig.1, it is found that a long $L_0$ needs that both the spin correlation and the charge correlation are strong. The $T_c^{max} - L_0$ relation is shown in Fig.2 for various superconductors–that is, the highest-$T_c$ superconductor has the largest $L_0$. However, because of the competitions between the charge correlation and the spin correlation, the size of $L_0$ is limited.

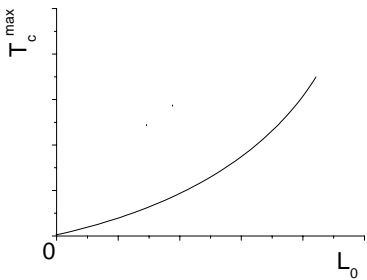

**Figure 2** The highest transition temperatures change with moderate correlation lengths for various superconductors.

Other factors also affect $T_c^{max}$ of superconductors, two materials having the same $L_0$ may have different $T_c^{max}$. However, $L_0$ dominate $T_c^{max}$ for the strong correlations induced superconductivity, no matter whether the superconductor has disorders.

Now let us explain the physics (1)-(9) in the ways how a material is turned into a superconductor. The ways of turning a material into a superconductor are to mediate both the spin correlation and the charge correlation to an appropriate level. Particularly, some materials may be mediated to have large $L_0$. All solid materials in normal states can be divided into such four categories: strong spin correlation materials (SCM), strong charge correlation materials (CCM), strong spin and charge correlation materials (SCCM), and weak correlation materials (WCM). For example, most lightly doped cupretas are SCM, most optimally doped cuprates are SCCM, and most heavily overdoped cuprates are CCM.

SCM in normal states usually are magnetic materials. To turn them into superconductors, we have to strengthen the charge correlations or weaken the spin correlations in them. One, by substituting non-magnetic elements for some magnetic elements in them, because the changes of the charge correlation with disorders are not monotonous, when the effects of charge correlations overcome the effects of disorders, they become superconductors. Two, by doping, like the cases in the p-type cuprate superconductors. Three, by extra influences, usual way to increase the charge correlation is to increase pressures. There are other ways. One of the ways is changing temperature. In most underdoped cuprates, the spin correlation decreases with decreased temperature, the charge correlation increases with decreased temperature, and this is beneficial to getting into superconducting state. While in most extremely lightly doped cuprates, the spin correlation increases with decreased temperature, the charge correlation decreases with decreased temperature, and this lead to the antiferromagnetic long-range order. Therefore, the changes of these correlations depend on the electronic structure and the occupying of electrons in energy band. The examples also include the transition from superconductor to ferromagnetism with decreased temperature in $ErRh_4B$[3], because the spin correlation increases with decreased temperature but the charge correlation decreases with decreased temperature in this material, the superconducting correlation is broken by the spin correlation at lower temperature on the basis of the physics(1). For the transition from antiferromagnetism to superconductor in $Upt_3$[4], because the increase of the charge correlation overcomes the change of the spin correlation with decreased temperature, there are two superconducting transition temperatures, may show the transition from magnetic excitations induced superconductivity to phonons induced superconductivity in my viewpoint. If one questioned my predictions, one had to exam the pairing symmetry or the isotope effect in the superconductivity of $Upt_3$. The differences between the charge correlations in $Upt_3$ and $ErRh_4B$ should be explained on their energy bands. Another way is with illumination. The high temperature superconductivity is usually broken by illumination in experiments, but I predict some lightly doped cuprates may show superconductivity in low temperature under moderate illumination. Moreover, other ways should include adding appropriate extra magnetic fields. Superconductivity is usually broken by magnetic field, since the magnetic field affects the spin correlation; I believe few magnetic materials may show superconductivity in weak extra field.

One particular example of SCM is NCO-type insulators. Combined with the ideas in another article[5], we can clarify why n-type superconductors have different properties from p-type superconductors. $Nd_{2-x}Ce_xCuO_4$ is an example[6,7]. The insulator $Nd_2CuO_4$ is SCM, but the charge correlation is quickly raised with increased $x$ in $Nd_{2-x}Ce_xCuO_4$, so that the material becomes the so-call SCCM around optimally doped region. As shown in Fig.3, although the highest transition temperature of the strong correlations induced superconductivity should appear at $x_1$, because the $L_0$ are not enough long in $Nd_{2-x}Ce_xCuO_4$ superconductors, the cooperation between magnetic excitations and phonons lead the highest-$T_c$ to appearing at $x_{opt}$. In this doping the charge correlation is stronger than the spin correlation, therefore, the resistivity behaves as $R \sim T^2$ in normal states, and both s-wave symmetry (from phonons) and d-wave symmetry (from strong correlations) coexist in superconducting states.

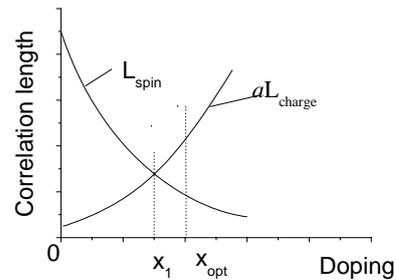

**Figure 3** The strong correlation induced highest-$T_c$ should appears at $x_1$, but the actual highest-$T_c$ appear at $x_{opt}$ due to the effects of phonons.

As shown in Fig.3, because the charge correlation is stronger than the spin correlation, both the spin excitations induced pairs and the phonons induced pairs may contribute to superconductivity at the same temperature. Moreover, on the basis of the competition mechanism, we can deduce that if the changes of the correlations with doping are quick, the pseudogaps is hard to be observed and the superconducting area to be narrow as shown in Fig.4. The solid curves '1' correspond to the case in $Nd_{2-x}Ce_xCuO_4$.

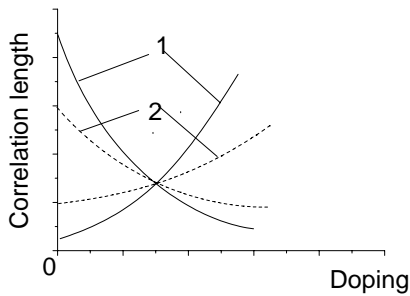

**Figure 2** The solid curves '1' correspond to the superconductors in which pseudogaps can be observed only when the strong correlations are very strong, while the dash curves '2' correspond to the superconductors in which pseudogaps are easy to be observed.

The very little isotope index of this superconductor is related to the physics (7).

CCM are usually phonons induced superconductors. In the past, one attributed some of them to the weak correlation materials. The known examples have overdoped cuprate superconductors and BCS-superconductors. Their resistance behaves as $R \sim T^2$ in normal states at low temperature as the feature (9) presented above, the s-wave symmetry (from phonons) dominates their superconductivity, and their isotope indices are usually near the normal value. $MgB_2$ is another example having these properties[8,9,10,11]. Because the strong charge correlation is beneficent to the superconducting correlation, the pairing temperature is near the superconducting transition temperature, as shown in BCS superconductors. Being contrary to SCM, to raise $T_c$, we have to strengthen the spin correlations or weaken charge correlation in them. For example, turning over-doped to optimally-doped cuprate superconductors is to strengthen the spin correlation, and making some materials into films is to weaken the charge correlation.

WCM are usually non-superconductors, such as good conductors Cu, Au, and Ag. That is to say, the charge correlation is a key factor to the superconducting correlation, while the spin correlation is a key factor to the electron pairing. To turn WCM into superconductors, we have to strengthen both the spin correlation and the charge correlation, this is a difficult problem. But this may be realized by substituting magnetic elements for some non-magnetic elements, or they must be bound into compounds with other elements. These ideas are consistent with experiments.

In summary, on the basis of the physics (1)-(9), we can predict the possible changes of the properties of solid materials qualitatively, and some predictions are testified in practice. Although the relations between various physics within materials are complex, the laws in them are simple.